\documentclass[fleqn,12pt,twoside]{article}
\usepackage{espcrc1}

\usepackage{graphicx}
\usepackage{epsfig}
\usepackage[figuresright]{rotating}

\newcommand{\AmS}{{\protect\the\textfont2
  A\kern-.1667em\lower.5ex\hbox{M}\kern-.125emS}}

\newcommand{\be}{\begin{equation}}
\newcommand{\ee}{\end{equation}}
\newcommand{\bea}{\begin{eqnarray}}
\newcommand{\eea}{\end{eqnarray}}
\newcommand{\f}{\frac}

\newcommand{\da}{\downarrow}

\newcommand{\Ga}{\Gamma}
\newcommand{\ga}{\gamma}

\newcommand{\ket}{\rangle}

\newcommand{\av}{{\rm{av}}}
\newcommand{\fl}{{\rm{fl}}}
\newcommand{\ina}{{\rm{in}}}

\newcommand{\ov}{\overline}

\title{Variance of the decay intensity of superdeformed 
bands\thanks{Supported by FAPESP.}} 
\author{M.~S.~Hussein\address[USP]
{Nuclear Theory and Elementary Particle Phenomenology 
Group, Instituto de F\'\i sica, Universidade de S\~{a}o Paulo,
Caixa Postal 66318, 05315-970 S\~{a}o Paulo, SP, Brazil}, 
A.~J.~Sargeant\addressmark[USP],
M.~P.~Pato\addressmark[USP] 
and 
M.~Ueda\address[ANCT]{Akita National College of Technology, 
Iijima Bunkyo-cho 1-1, Akita, 011-8511, Japan}}
\begin{document}
\maketitle

\begin{abstract}
We present analytic formulae for the energy average and variance of the 
intraband decay intensity of a superdeformed band. 

\end{abstract}

\section{Introduction}

The intensity of the collective $\ga$--rays emitted during the cascade down
a superdeformed (SD) band remains constant until a certain spin is reached 
where-after it drops to zero within a few transitions. 
The sharp drop in intensity is believed to arise from mixing of the SD states 
with normally deformed (ND) states of identical spin \cite{Vigezzi:1990a}. 
The model of Refs.~\cite{Shimizu:1993,Yoshida:2001} 
attributes the suddenness of the decay-out to the spin dependence of the 
barrier separating the SD and ND minima of the deformation potential.
Refs. \cite{Aberg:1999a,Krucken:2000,Sargeant:2001aq} discuss the effect 
of the chaoticity of the ND states on the decay-out.

In the present paper we present analytic formulae for the energy average 
(including the energy average of the fluctuation contribution) and variance 
of the intraband decay intensity of a superdeformed band
in terms of variables which usefully describe the decay-out
\cite{Krucken:2001we,Gu:1999bv,Sargeant:2002sv}. 
In agreement with Gu and Weidenm\"uller \cite{Gu:1999bv} (GW)
we find that average of the total intraband decay
intensity can be written as a function of the dimensionless variables
$\Ga^\da/\Ga_S$ and $\Ga_N/d$ where
$\Ga^\da$ is the spreading width for the 
mixing of an SD state with ND states of the
same spin, $d$ is the mean level spacing of the latter and 
$\Ga_S$ $(\Ga_N)$ are the electromagnetic decay widths of the SD (ND) 
states. Our formula for the variance of the total intraband decay intensity,
in addition to the two dimensionless variables just mentioned,
depends on the dimensionless variable $\Ga_N/(\Ga_S+\Ga^\da)$. 

\section{Energy average and variance of the decay intensity}

The total intraband decay intensity has the form 
\cite{Sargeant:2001aq,Gu:1999bv}
\be
I_\ina=\left(2\pi\Ga_S\right)^{-1}\int_{-\infty}^{\infty}
dE|A_{00}(E)|^2,
\label{I}
\ee
where $A_{00}(E)$ is the intraband decay amplitude 
and $\Ga_S$ is the electromagnetic decay width of 
superdeformed state $|0\ket$.
The energy average of Eq.~(\ref{I}) may be written 
as the incoherent sum \cite{Sargeant:2002sv,Kawai:1973}
\be
\ov{I_\ina}=I_\ina^\av+\ov{I_\ina^\fl},
\label{ovI}
\ee
where
\be
I_\ina^\av=\ov{I_\ina^\av}=\left(2\pi\Ga_S\right)^{-1}
\int_{-\infty}^{\infty}dE
\left|\ov{A_{00}(E)}\right|^2
\label{Iav}
\ee
and
\be
\ov{I_\ina^\fl}=\left(2\pi\Ga_S\right)^{-1}
\int_{-\infty}^{\infty}dE
\ov{\left|A_{00}^\fl(E)\right|^2},
\label{ovIfl}
\ee
where we have written the decay amplitude as
$A_{00}=\ov{A_{00}}+A_{00}^\fl$ where $\ov{A_{00}}$ is a background 
contribution and $A_{00}^\fl$ is the fluctuation on that background.
In \cite{Sargeant:2002sv} the background is taken to be
\be\label{2Aav}
\ov{A_{00}}=\f{\Ga_S}{E-E_0+i(\Ga_S+\Ga^\da)/2}.
\ee
Eq.~(\ref{2Aav}) exhibits the structure of an isolated
doorway resonance. The doorway $|0\ket$ has an escape width $\Ga_S$ for 
decay to the SD state with next lower spin and a spreading width $\Ga^\da$ for
decay to the ND states with the same spin which are reached
by tunnelling through the barrier separating the SD and ND wells.

In \cite{Sargeant:2002sv} it is shown that the auto-correlation function of 
the decay amplitude is given by
\bea
\ov{A_{00}^\fl(E){A_{00}^\fl(E')}^*}
&\approx&2\left(2\pi\Ga_N/d\right)^{-1}
(\Ga^\da/\Ga_S)^2\hspace{2mm}
{\ov{A_{00}(E)}}^2\hspace{2mm}\f{i\Ga_N}{E-E'+i\Ga_N}\ov{{A_{00}(E')}^*}^2.
\label{2Aflauto}
\eea
When $E'=E$ this reduces to  
\be
\ov{\left|A_{00}^\fl\right|^2}=2\left(2\pi\Ga_N/d\right)^{-1}
\f{{\Ga_S}^2{\Ga^\da}^2}
{\left[(E-E_0)^2+(\Ga_S+\Ga^\da)^2/4\right]^2},
\label{Aflsqu}
\ee
which is the average of the fluctuation contribution to the transition 
intensity.

The integrals in Eqs.~(\ref{Iav}) and (\ref{ovIfl}) may be carried out
using the calculus of residues. One obtains 
\be\label{2Iav}
I_\ina^\av=\f{1}{1+\Ga^\da/\Ga_S},
\ee
for the average background contribution and
\bea
\ov{I_\ina^\fl}&=&2\left(\pi\Ga_N/d\right)^{-1}
\f{\left(\Ga^\da/\Ga_S\right)^2}
{\left(1+\Ga^\da/\Ga_S\right)^3}
\label{2ovIfl}
=2\left(\pi\Ga_N/d\right)^{-1}
I_\ina^\av\left(1-I_\ina^\av\right)^2,
\label{3ovIfl}
\eea
for the average fluctuation contribution to the average decay intensity.
\begin{figure}
\begin{minipage}[t]{0.48\linewidth} 
\centering
\includegraphics[width=.95\linewidth]{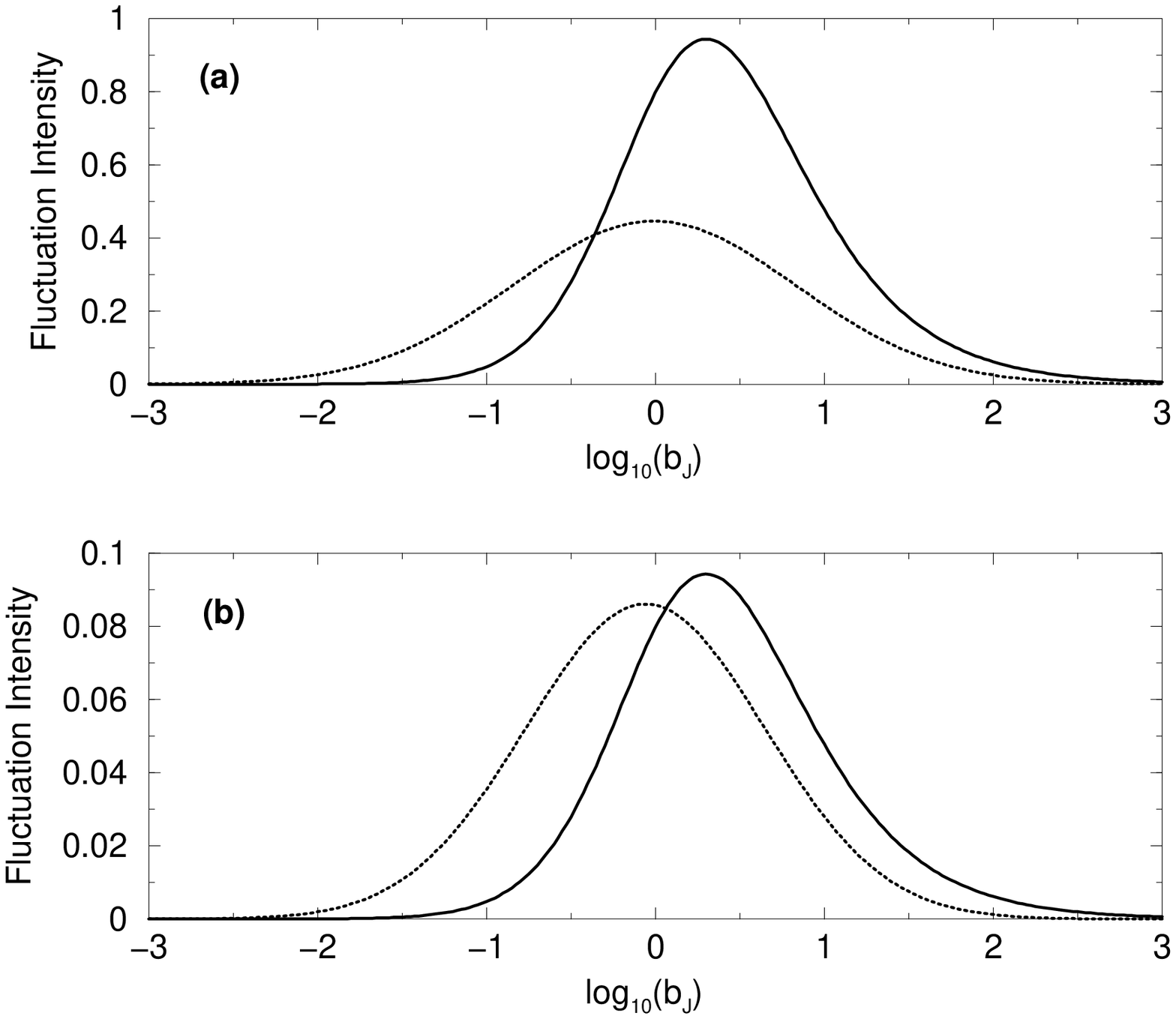}
\label{flucfig1}
\caption{Average of the fluctuation contribution to the intraband intensity 
$\ov{I_\ina^\fl}$~vs.~$\log_{10}(b_J)$ where $b_J\equiv\Ga^\da/\Ga_S$. 
The solid 
lines were calculated using Eq.~(\ref{2ovIfl}) and the dotted lines by GW's 
fit formula, Eq.~(\ref{gufit}). The variable $\Ga_N/d$ took the value 0.1 for
graph (a) and 1 for graph (b).}
\end{minipage}
\hspace{0.02\linewidth} 
\begin{minipage}[t]{.48\linewidth}
\centering
\includegraphics[width=.95\linewidth]{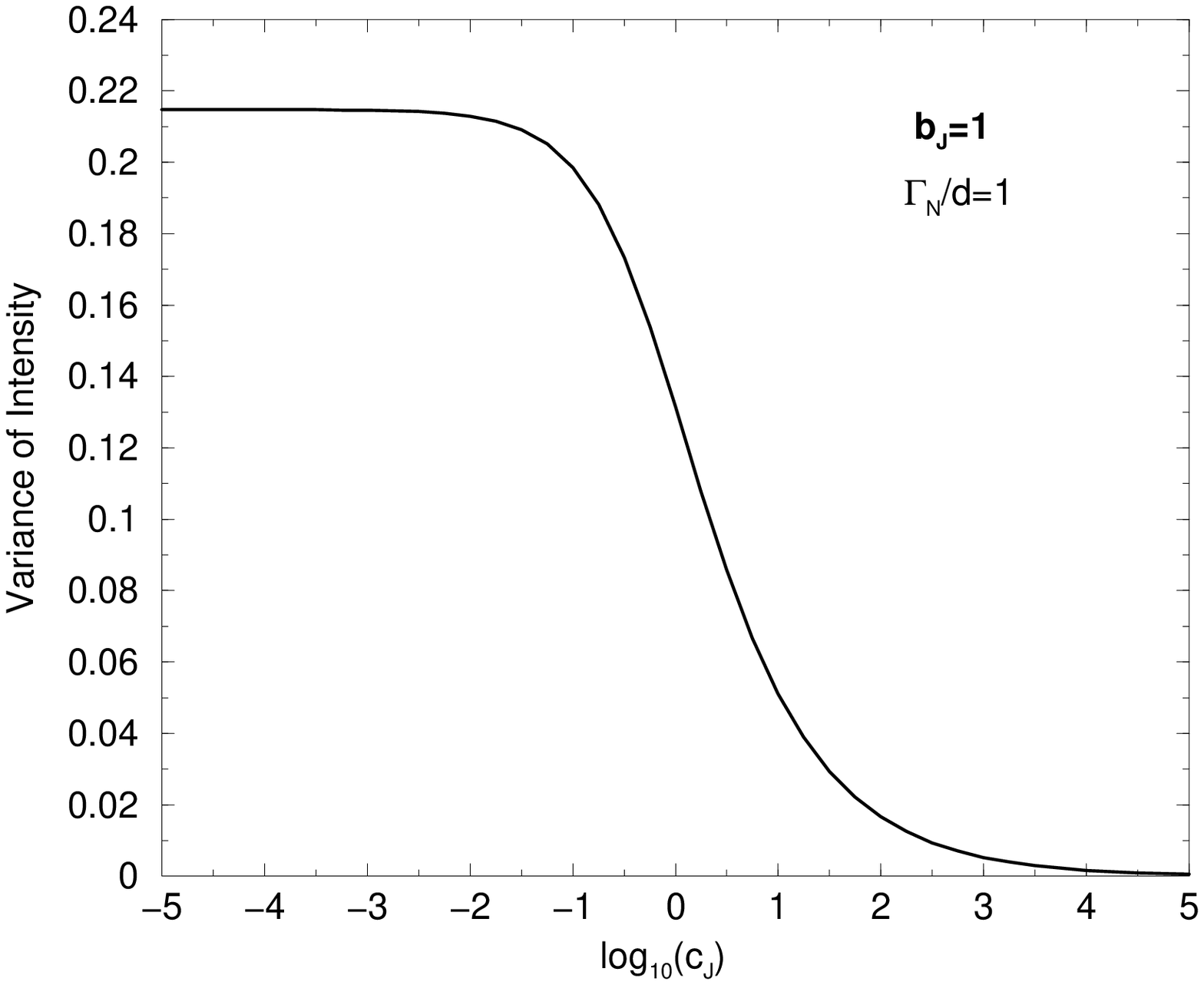}
\label{flucfig4}
\caption{The standard deviation of the decay intensity 
$\sqrt{\ov{\left(\Delta I_\ina\right)^2}}$~vs.~$\log_{10}(c_J)$ 
where $c_J\equiv \Ga^\da/\Ga_N$ plotted using Eq.~(\ref{3var})
for fixed $b_J=\Ga^\da/\Ga_S$ and $\Ga_N/d$.}
\end{minipage}
\end{figure}
Eq.~(\ref{2ovIfl}) for $\ov{I_\ina^\fl}$ is plotted in Fig.~\ref{flucfig1} 
and for comparison we have also plotted a fit formula which was 
obtained by GW,
\bea
\ov{I_\ina^\fl}&=&\left[1-0.9139\left(\Ga_N/d\right)^{0.2172}
\right]
\exp\left\{-\f{\left[0.4343\ln\left(\f{\Ga^\da}{\Ga_S}\right)
-0.45\left(\f{\Ga_N}{d}\right)^{-0.1303}\right]^2}
{\left(\Ga_N/d\right)^{-0.1477}}\right\}.
\label{gufit}
\eea
Qualitative agreement is seen between the two formulae.  
Our results are strictly valid only when $\Ga_N/d\gg 1$.
The dependence of $\ov{I_\ina^\fl}$ (and that of $I_\ina^\av$) 
on $\Ga^\da/\Ga_S$ results from the resonant doorway energy 
dependence of the decay amplitude $\ov{A_{00}(E)}$ [Eq.~(\ref{2Aav})]. 
This energy dependence also manifests itself in the average of the fluctuation
contribution to the transition intensity $\ov{|A^\fl_{00}(E)|^2}$ 
[Eq.~(\ref{Aflsqu})]. GW include precisely the same energy dependence
in their calculation by use of an energy dependent transmission coefficient to 
describe decay to the SD band.
This is the reason for our qualitative agreement with
GW concerning $\ov{I_\ina}$. 

A measure of the dispersion of the calculated $I_\ina$ is given by the
variance 
\be\label{var}
\ov{\left(\Delta I_\ina\right)^2}
=\ov{\left(I_\ina-\ov{I_\ina}\right)^2}.
\ee
It is shown in \cite{Sargeant:2002sv} that
\be\label{3var}
\ov{\left(\Delta I_\ina\right)^2}=
{\ov{I_\ina^\fl}}^2f_1\left(\xi\right)
+2I_\ina^\av\ov{I_\ina^\fl}f_2\left(\xi\right),
\ee
where the variable $\xi$ is defined by
\bea\label{xi}
\xi\equiv\f{\Ga_S+\Ga^\da}{\Ga_N}
=\f{\Ga_S}{\Ga_N}(1+\Ga^\da/\Ga_S)=\f{\Ga_S}{\Ga_N}
{I_\ina^\av}^{-1}
=\f{\Ga^\da}{\Ga_N}(1+\Ga_S/\Ga^\da)^{-1}
=\f{\Ga^\da}{\Ga_N}(1-I_\ina^\av).
\eea
and
\be
f_1\left(\xi\right)=\f{1}{\left(1+\xi\right)}+\f{\xi}{\left(1+\xi\right)^2}
+\f{\xi^2}{2\left(1+\xi\right)^3}
\hspace{.5cm}\mbox{and}\hspace{.5cm}
f_2\left(\xi\right)=\f{1}{2\left(1+\xi\right)}.
\ee

Since the variance depends only on $(\Ga_S+\Ga^\da)/\Ga_N$ in addition to
$\Ga^\da/\Ga_S$ and $\Ga_N/d$, upon fixing the latter two variables
the variance may be considered a function of any {\em one} of $\Ga^\da/\Ga_N$,
$\Ga_S/\Ga_N$, $\Ga^\da/d$ or $\Ga_S/d$ [see Eq.~(\ref{xi})]. 
Fig.~\ref{flucfig4} shows a plot of the standard deviation,
$\sqrt{\ov{\left(\Delta I_\ina\right)^2}}$, [Eq.~(\ref{3var})] as a 
function of $\Ga^\da/\Ga_N$ for fixed $\Ga^\da/\Ga_S$ and $\Ga_N/d$.
Ultimately, the variance like the intensity is a function of the spin of the
decaying nucleus and could provide an additional probe to the spin 
dependence of the barrier separating the SD and ND wells which is contained
in the spreading width $\Ga^\da$ 
\cite{Vigezzi:1990a,Shimizu:1993,Yoshida:2001}. 
Our result for the variance of the decay intensity, 
$\ov{\left(\Delta I_\ina\right)^2}$ [Eq.~(\ref{3var})] has a structure 
reminiscent of Ericson's expression for the variance of the cross section 
\cite{Ericson:1963}. 
In the case compound nucleus scattering, extraction of the correlation
width from a measurement of cross section autocorrelation function
permits the determination of the density of states of the compound nucleus
\cite{Ericson:1966}. In the present case the variance supplies a second ``equation'' 
besides that for $\ov{I_\ina}$.
Both equations are functions of $\Ga^\da$ and $d$, since the electromagnetic
widths are measured. Thus both $\Ga^\da$ and $d$ can be unambiguously 
determined.
\section{Conclusions}
In conclusion, we have presented analytic formulae for the energy average 
and variance of the intraband decay intensity of a superdeformed band. 
These formulae were 
derived by making assumptions and approximations which are strictly valid 
only in the strongly overlapping resonance region, $\Ga_N/d\gg 1$. 
However, these formulae work 
well when $\Ga_N/d$=1 and provide a qualitative description even when 
$\Ga_N/d$=0.1.
We have revealed that the variance of the decay intensity depends on the correlation 
length $\Ga_N/(\Ga_S+\Ga^\da)$ in addition to the two dimensionless variables
$\Ga^\da/\Ga_S$ and $\Ga_N/d$ on which the average of the decay intensity
depends. 

\bibliography{sargeant,rmt,sd,compound}
\bibliographystyle{npa}

\end{document}